\documentclass[conference]{IEEEtran}
\IEEEoverridecommandlockouts
\usepackage{cite}
\usepackage{amsmath,amssymb,amsfonts}
\usepackage{algorithmic}
\usepackage{graphicx}
\usepackage{textcomp}
\usepackage{tabularx}
\usepackage{multicol}
\def\BibTeX{{\rm B\kern-.05em{\sc i\kern-.025em b}\kern-.08em
    T\kern-.1667em\lower.7ex\hbox{E}\kern-.125emX}}
\begin{document}

\onecolumn
\textcopyright 2017 IEEE. Personal use of this material is permitted. Permission from IEEE must be obtained
for all other uses, in any current or future media, including reprinting/republishing this material for advertising or promotional purposes, creating new collective works, for resale or redistribution to servers or lists, or reuse of any copyrighted component of this work in other works.
\newline
\newline
The final, published version of this paper is available under:
S. Marksteiner, V. J. Exp\'osito Jim\'enez, H. Vallant, and H. Zeiner, "An Overview of Wireless IoT Protocol Security in
the Smart Home Domain," \textit{2017 Joint 13th CTTE and 10th CMI Conference on Internet of Things Business Models,
Users, and Networks}, Copenhagen, 2017, pp. 1-8. doi: 10.1109/CTTE.2017.8260940.
http://ieeexplore.ieee.org/document/8260940/
\newpage
\twocolumn

\title{An Overview of Wireless IoT Protocol Security in the Smart Home Domain
}

\author{\IEEEauthorblockN{Stefan Marksteiner, V\'ictor Juan Exp\'osito Jim\'enez, Heribert Vallant, Herwig Zeiner}
\IEEEauthorblockA{
	JOANNEUM RESEARCH Forschungsgesellschaft mbH\\
	DIGITAL - Institute for Information and Communication Technologies\\
	Graz, Austria\\ 
	Email: stefan.marksteiner@joanneum.at
	}

}

\maketitle

\begin{abstract}
While the application of IoT in smart technologies becomes more and more proliferated,
the pandemonium of its protocols becomes increasingly confusing. More seriously, severe security deficiencies of these
protocols become evident, as time-to-market is a key factor, which satisfaction comes at the price of a less thorough
security design and testing. This applies especially to the smart home domain, where the consumer-driven market demands
quick and cheap solutions.
 This paper presents an overview of IoT application domains and discusses the most important wireless IoT
protocols for smart home, which are KNX-RF, EnOcean, Zigbee, Z-Wave and Thread. Finally, it describes the security
features of said protocols and compares them with each other, giving advice on whose protocols are more suitable for
a secure smart home.
\end{abstract}


\section{Introduction}
\subsection{Motivation}
Although wireless sensor connections offer several ways to increase our productivity in many fields such as smart home,
smart production or smart transportation, it also introduces some risks to be aware of. The usage of a wireless physical
communication, which allows attackers easier interception of communications, together with the Internet of Things
(IoT)~\cite{hyperconnected2017}~\cite{7879243} or Web of Things (WoT)~\cite{zeiner2016} also leads to unprecedented
opportunities for attackers to reveal confidential information and to manipulate data. It is crucial to find efficient
and effective methods to counteract such attacks. Otherwise, all the benefits of the IoT will be forfeit.

In order to address these challenges, first, a deep security analysis of the existing technologies is needed to help
discover the root causes as well as find analysis techniques that allow verifying the security of the system. Moreover,
other aspects also have to be considered to reach a secure environment. In some cases, there are no resources to
implement the needed secure methods, for example, on sensor nodes with limited resources that operate in adverse
environments in which very efficient methods have to be provided. On the other hand, security is not only a hardware
method. For this reason, software attacks, for example attacks against memory consumption, have to be always in scope to
avoid them.

An analysis about the security on the IoT would be huge and it can not be done just in one publication. Because of this
fact, our research is focused on a security analysis of the main wireless protocols in the smart home domain. This
publication has three main parts. The next section gives an overview of the different domains that can be found in the
Internet of Things. Section~\ref{sec:tec} describes the way in which the sensors can be connected as well as a brief
presentation of the selected wireless protocols. The subsequent Section~\ref{sec:security} provides a security
analysis of each protocol and Section~\ref{sec:conclusion}, eventually, shows the conclusion and outlook of our
research.

\subsection{Related Work}
Security and privacy are not simple tasks and include several different issues to carry out in an IoT domain, the
article in ~\cite{7867732} gives us an overview of the most common challenges in this field. On the other hand, Granjal
et al.~\cite{7005393} present an exhaustive analysis of the security and privacy of each layer of the OSI model
according to the existing protocols and their implication in the general IoT domain. Focusing on the smart home domain
topic, an extended analysis of the security is detailed in~\cite{Batalla:2017:SSH:3145473.3122816}. It contains an
in-depth report of the main aspects of this area, such as the most common threats and good practices as well as a brief
protocol analysis and security implications of using of cloud platform on smart homes. One step beyond is given
from~\cite{7930376}, in which the authors not only describe the main security and privacy threads, but introduce an
algorithm to secure each situation. Finally, they test them in a real-environment with successful results. Although, it
seems that these issues only concern researchers, the conclusions of the research~\cite{Brush:2011:HAW:1978942.1979249}
present that one of the main problems customers find to implement smart home solutions is security followed by
inflexibility, costs, and poor manageability, which indicates that it is also an important point for customers.

Unfortunately, sometimes the theoretical research is not enough and a practical research is needed. For
example,~\cite{7321811} introduces a study about the Google Nest\footnote{https://nest.com/} and the Nike+
Fuelband\footnote{https://www.nike.com/nike-plus}, in which both hardware and software are analyzed. Another interesting
approach is given in~\cite{7397713}, in which the security and privacy are tested in different IoT demos, such as a
small light system. It describes detailed analysis and depicts the possible risks for each scenario.
Moreover, a new kind of devices to take into account are the low energy devices in which a security system has to be
applied but is constrained by an extremely low power usage as the article~\cite{7031838}
describes.~\cite{Sivaraman:2016:SAS:2939918.2939925} shows why a smart home scenario cannot be considered as an isolated system by proving it can be exploited by using an external mobile application.

Along with this related work, we have seen numerous security and privacy threads and also some possible solutions, but
not an extended security comparison of the available protocols that can be used in the smart home domain, which is the
main focus of this paper.

\section{IoT Application Domains}
\label{sec:domains}
\begin{figure}[ht]
	\centering
		\includegraphics[width=1.0\linewidth]{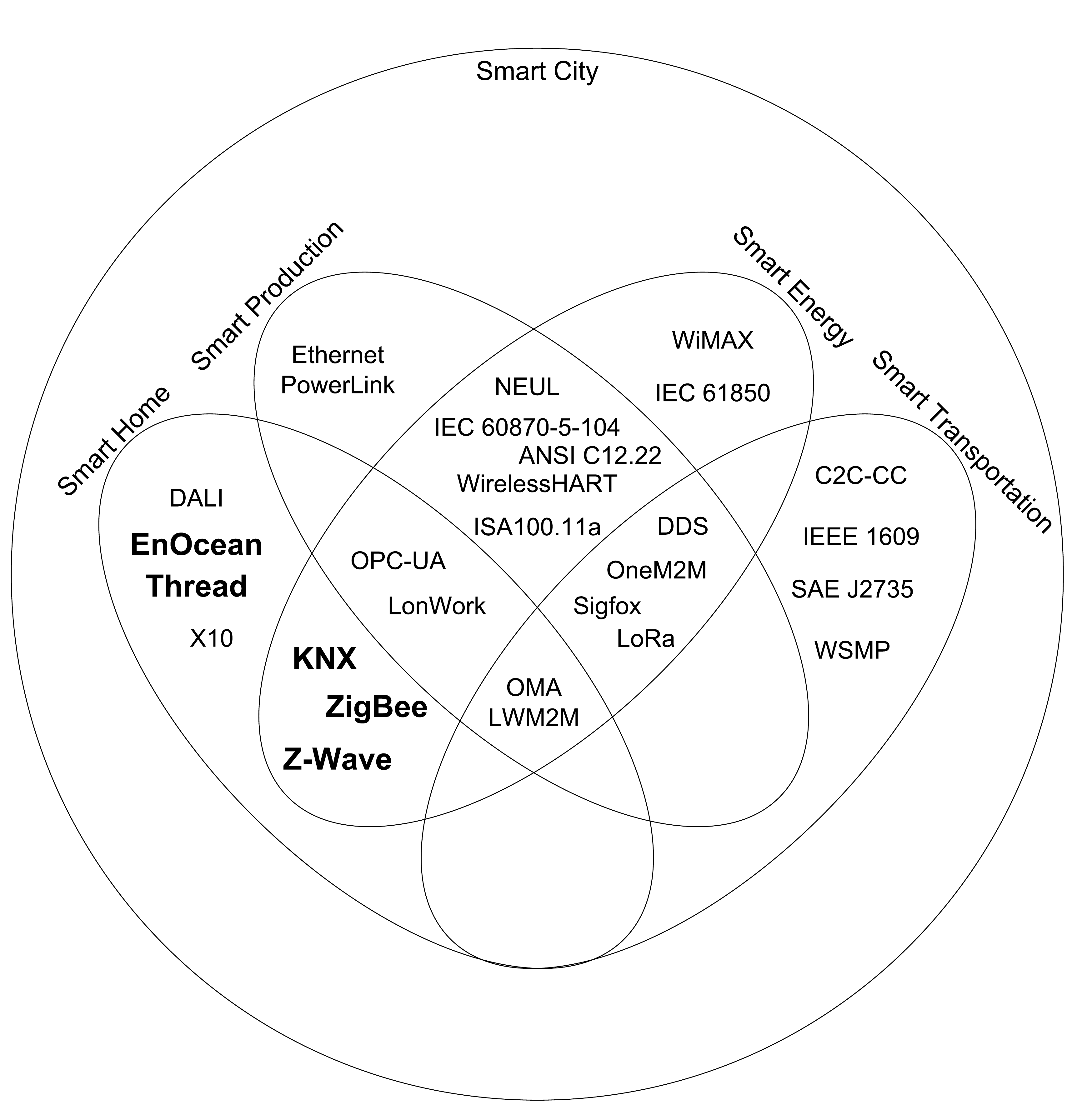}
		\caption{Venn diagram of IoT application domains and included protocols}
	\label{fig:smart_dom}
\end{figure}

Figure \ref{fig:smart_dom} displays an overview of the (smart) IoT application domains considered in this paper,
including the communication protocols used in each respective area.

\subsection{Smart Home} 
The smart home market is getting more and more dynamic and, according a Smart Home Customer Survey of
Deloitte~\cite{deloitte2015},
in 2018 one million households could already be smart in Germany. According to that study, the main interests in Smart
Home are closely linked to more comfort and safety, followed by savings on heating and electricity costs. The main
barriers for customers are on the one hand the costs and on the other the concerns regarding data protection and data
security. One important recommendation for the market players is to address security and privacy adequately and make it
transparent to the customer. Within this paper we give an overview of the most common wireless protocols used in the
Smart Home domain and an analysis of the defined security measures.

\subsection{Smart Production}
 Recent reports~\cite{zhang2017}~\cite{wang2016}, describe key issues for the next generation of smart production
 analytic services. Relevant applications are: digital performance management (including a data-driven mindset and
 integration across previously isolated functions); predictive maintenance (including integration of diverse data sets
 and using, e.g., advanced self-learning algorithms); yield, energy, and throughput optimization (including integration
 of process control with other data); next-level automation (including improvements in sensor technology and demand
planning); and digital quality management (including the use of new sensing technologies and semi-automated quality
control). In Smart Production, wireless sensor networks will play a key role for increasing the flexibility of a data
driven production lifecycle. Furthermore, for such a connected environment, it is clear that we should deal with cyber
security issues described in this paper.

\subsection{Smart Transportation} 
Smart transportation is becoming one of the biggest domains of the IoT. The implementation of the Controller Area
Network (CAN)~\cite{can2013}, that is commonly used in the automation control together with new protocols and
communication technologies such as the 5G~\cite{7169508} or IoT-Narrow Band (IoT-NB)~\cite{7925809}, opens new
possibilities to exchange information. These new technologies are able  to give smart transportation the necessary
packet delay and data transmission rate. Moreover, new hardware implementations, specifically designed to make the right
decision as fast as possible, provide a new key tool for the future of the autonomous vehicles. In this future they will
have to be able to not only communicate with other cars or services, but process all information of the environment in real
time to make the right choice. In consequence, the protection of all sensible information, as well as communications, has to be
a mandatory point to ensure the safety and privacy of users.

\subsection{Smart Energy} 
In the energy domain, several standards are available for different areas ranging from generation, transmission,
distribution and distributed energy resources  to the customers, which may be also producers themselves, making them
so-called prosumers.
A good overview of these standards is available at the International Electrotechnical Commission
Website\footnote{http://smartgridstandardsmap.com/}. Regarding communication networks, the whole range beginning from
home area networks, located at the customer, over the field area networks at the distribution level and wide area
networks at transmission level are represented. That means wireless standards like 2G/3G/4G, WiMAX,
WLAN, WirelessHART, ISA100.11a, ZigBee, Z-Wave, 6LoWPAN, LoraWan, Sigfox, as well as wired standards such as Ethernet,
profibus, profinet, modbus or PLC are used  for connectivity.
Frequently, smart energy gateways are used to consolidate communications, having their
own security requirements \cite{Marksteiner:2016:7973855}.

\section{Wireless Smart Home Protocols}
\label{sec:tec}

Before going deep into the protocols in IoT, it is highly recommended to start on Open Systems Interconnection (OSI)
layer model which gives a better understanding of the implementation of the explained protocols. The OSI layer
model~\cite{itu200} provides a standard architecture to define network communication with the following layers:

\begin{itemize}
\item\textit{Layer 1 - Physical:} Information of bits through a physical medium;
\item\textit{Layer 2 - Data link:} Controls errors in transmission between two adjacent nodes by using frames;
\item\textit{Layer 3 - Network:} Adds the concept of routing in which a frame is able to reach a destination beyond adjacent nodes;
\item\textit{Layer 4 - Transport:} Reliable transmission of data, in which new optional capabilities can be added such as retransmission or flow control;
\item\textit{Layer 5 - Session:} Manages the sequence and flow of events;
\item\textit{Layer 6 - Presentation:} Manages the syntax processing of messages to be used in the application layer such as the encryption/decryption;
\item\textit{Layer 7 - Application:} The layer where end-user applications are implemented.      
\end{itemize} 

\begin{figure}[ht]
	\centering
		\includegraphics[width=0.7\linewidth]{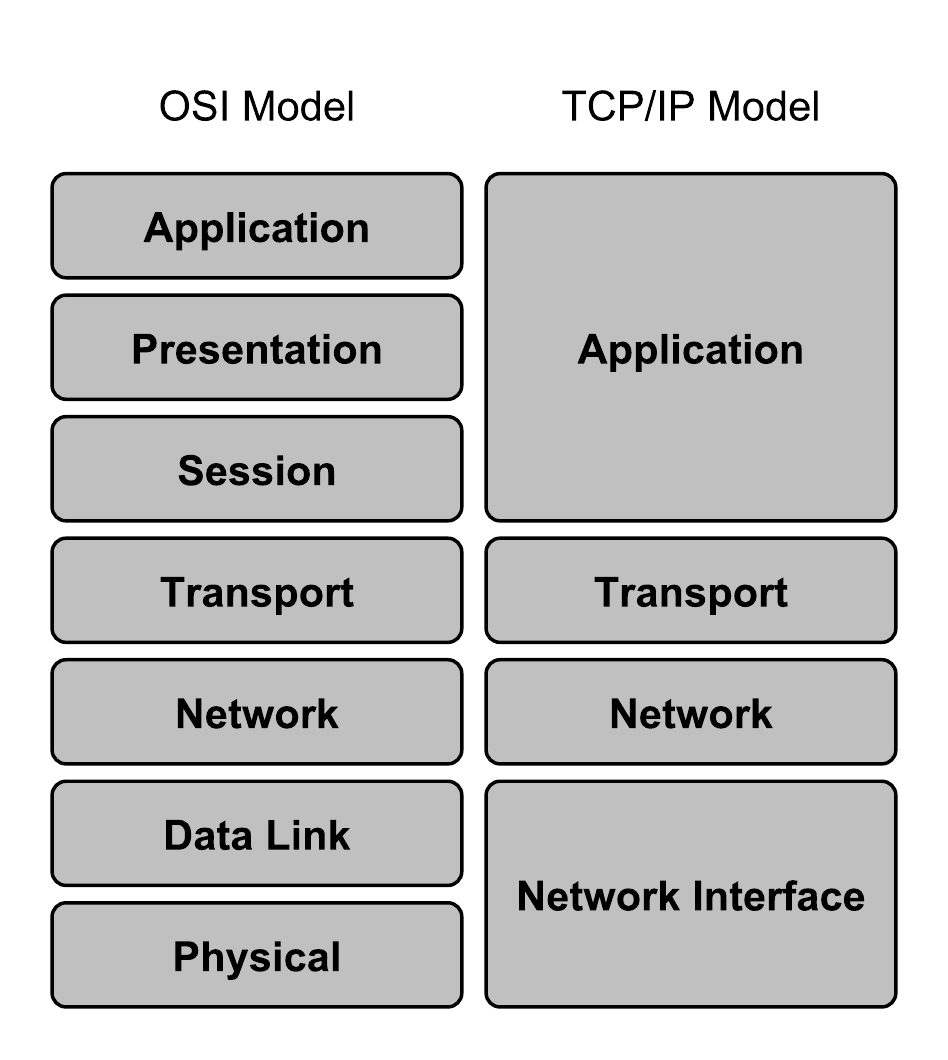}
		\caption{Overview of the OSI model on the different protocols}
	\label{fig:osi-model}
\end{figure}

Another approach is given in the TCP/IP model~\cite{rfc1122,rfc1123}, which is frequently used for internet
communications because it is a simplified vision of the OSI layer model by only using four layers. The
figure~\ref{fig:osi-model} shows a representation of both versions

In the field of the IoT, \cite{8012291} proposes an IoT stack as the combination of four main components: the physical layer, the IoT platform, the communication protocols/technologies and the application layer. Instead, \cite{5579493} suggests an architecture based on three main layers: the perception layer, the network layer, and the application layer. Following these approaches, this publication is focused on the network layer in which several protocols and communication technologies are described as well as the security and privacy aspects of each of the selected domains.

Another important point to consider is the way in which devices are connected to each other on a wireless domain. Although there are several topologies, two main types can be differentiated:
\begin{itemize}
	\item\textit{Centralized or star:} There is central node o hub which is the responsible one for managing communications with other nodes of the network and the outside;
	\item\textit{Decentralized or fully connected:} All nodes are connected to other network nodes; this kind of topology is not efficient when the network is bigger, because the communication effort grows exponential with the number of.
\end{itemize}

\begin{figure*}[ht]
	\centering
		\includegraphics[width=1.0\linewidth]{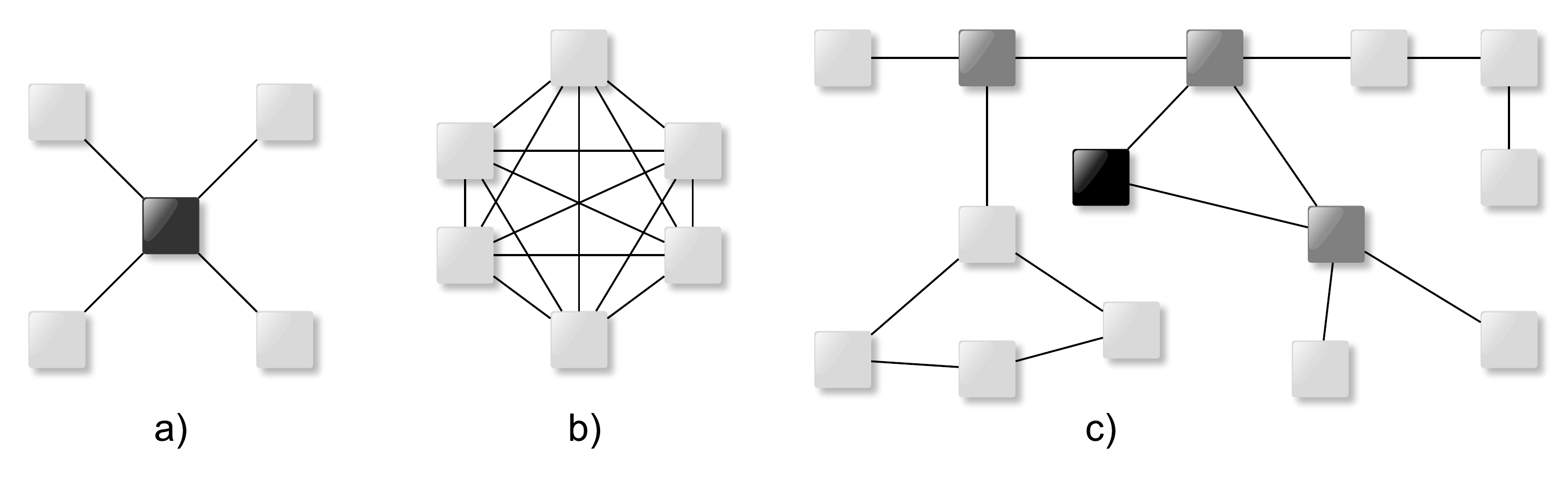}
		\caption{Network topologies in this publication: a) Star b) Fully connected c) Mesh}
	\label{fig:network-topologies}
\end{figure*}

A middle ground between the topologies above is the mesh topology, in which several nodes are able to communicate with
each other through the communication between intermediary nodes. The figure~\ref{fig:network-topologies} depicts an example of a mesh network, which also may include other topologies such as the following:
\begin{itemize} 
	\item\textit{Ring:} All nodes make a loop in which each node is connected to two nodes; the information goes through each node until it reaches the destination node;
	\item\textit{Bus:} All nodes are connected to a backbone and communications and messages go through it in which all
	nodes receive all messages; if the backbone cable fails, all networks fail;
	\item\textit{Line or point-to-point:} Is the simplest network topology, which consists of the connection between two 	endpoints.
\end{itemize} 

This research is focused on the wireless protocols which are available in the smart home domain. These requirements
reject the protocols: X10~\cite{X10spec}, DALI~\cite{4582728}\cite{6175666}, and LonWorks~\cite{lonworks2001} that
mainly use power line and twisted pair as on the physical layer. It is also focused on the security implications of
protocols on the network and transport layers of the OSI model. According to that, OMA Lightweight M2M (OMA
LWM2M)~\cite{7389115} and OPC Unified Architecture (OPC UA)~\cite{OPCUA:2017} were also discarded because are
implemented beyond the transport layer. Then, the chosen protocols are the following:

\subsection{KNX-RF}
\label{sec:tec:prt:knx}

Developed in 1991, KNX\footnote{https://www.knx.org/} is one of the main protocols in the Heating, Ventilation and Air
Conditioning (HVAC), with a huge number of compatible devices available on the market. Moreover, in 2006, it was
accepted as an open standard by the ISO/IEC 14543-3 specification. The protocol is based on the OSI model and covers the
data link, network, and transport layer. In this paper, we focus on KNX Radio Frequency (RF), although the protocol
is not limited to this physical medium and is able to use other transmission mediums such as the twisted pair, power
line, and ethernet (KNXnet/IP). The used band for the wireless transmission is located in the industrial, scientific and
medical (ISM) bands, specifically at 868 MHz and 2.4 GHz and is able to reach a maximum range up to 150 meters. The
maximum data rate transmission is up to 16.385 kbps.

The nodes are connected through tree topologies: line, tree, and star. A KNX network is based on areas and lines. Each
area can include a maximum of 15 lines where the devices are connected as end nodes. The maximum number of devices that
can be addressed is 65k, which can communicate with each other without any master device because KNX is a peer-to-peer system.

\subsection{EnOcean}
\label{sec:tec:prt:enocean}

Although it was patented in 2001, in 2012 it became an international standard (ISO/IEC 14543-3-10). Its main feature is
the wireless power supply which allows devices be independent of a battery to work, because they receive their energy
from the wireless signal. Th  EnOcean standard involves the three lowest layers of the OSI model; physical, data, and
network. To ensure a better integration, the EnOcean alliance provides the EnOcean Equipment Profiles (EEP) layer which
is located in the application layer to reach interoperability between different type of products and suppliers with this
standard. It uses the ISM bands as transmission frequency; 868 MHz, 315 MHz, and, since May 2017, 2.4 GHz through
Easyfit\footnote{https://www.easyfit-solutions.com/}. EnOcean uses a mesh topology in which all nodes communicate with
each other. The signal range is up to 300 meters in free field and 30 meters inside a building and the maximum data rate is 125 kpbs.

\subsection{Zigbee}
\label{sec:tec:prt:zigbee}

Zigbee was developed in 2001 and updated by the Zigbee PRO specification in 2007. The latter is fully backward
compatible and includes some improvements, such as a better security. Unique about Zigbee is the
implementation of concrete specifications for different scenarios. Two examples are Zigbee LightLink that is widely used
on smart light systems (i.e Philips Hue\footnote{http://www2.meethue.com/}), or Zigbee Green Power, which is able to
work with battery-less devices in a similar way EnOcean does. Zigbee can be used with three topologies; star, tree and
mesh in which a maximum of 65k nodes are supported. Zigbee operates in the ISM bands, 913 MHz, 868 MHz and 2.4 GHz and the
devices range is limited to 10-100 meters. The maximum provided data range is up to 20kbps within 913 MHz and 868 MHz
bands and 250 kbps for the 2.4 GHz band. Zigbee uses the IEEE 802.15.4 standard as a physical and data link layer.

\subsection{Z-Wave}
\label{sec:tec:prt:zwave}
Developed in 2001, Z-Wave is mainly focused on wireless lightweight and low-latency transmission data. The newest update
of the protocol, called Z-Wave Plus, was given in 2013 and adds some improvements such as better battery life as well as
wireless range. Unlike Zigbee or EnOcean, Z-Wave is not a standard and its development is controlled by the Z-Wave
Alliance\footnote{https://z-wavealliance.org/} which involves over 600 companies including big players of the IoT sector
like Siemens or Huawei. Z-Wave uses a mesh network with the number of connected devices in a network limited to up to
232 nodes. Z-Wave involves the four lower layers of the OSI models, physical, data link, network and transport.
Moreover, the implementation of the physical and data link layers has been included as standard G.9959 by the
International Telecommunication Union (ITU). It is able to work on the common industrial frequency 828 MHz at EU
markets and 908 MHz as a part of the ISM bands with a maximum range of up to 30 meters. Finally, the maximum provided
data rate is up to 100 kbps.

\subsection{Thread}
\label{sec:tec:prt:threadgroup}

Thread\footnote{https://threadgroup.org/} is a standard that has been specially designed for wireless device-to-device
communications. It can be used on small and large networks together with low-power devices. One of the strongest
advantages of the stack is that there is no single point of failure. If a slave device depends on a
master device that is unavailable, the child is able to independently select another master device. This is possible
because it uses a mesh network topology together with 6LoWPAN, in which every node can act as master node and there is
no limit to connected nodes due to the usage of IPv6. 6LowWPAN is also used in the version 4.2 of
Bluetooth\footnote{https://www.bluetooth.org} Low Energy (BLE) through the Internet Protocol Support
Profile\cite{RFC7668}, which enables Bluetooth Smart sensors to access the Internet directly via 6LoWPAN connectivity.
The Thread procotol implementation is given in the layers number 3 and 4 of the OSI layer model and uses the IEEE
802.15.4 standard in the layers 1 and 2, which allows a data rate of 250 kbps.

\begin{figure*}[ht]
	\centering
		\includegraphics[width=1.0\linewidth]{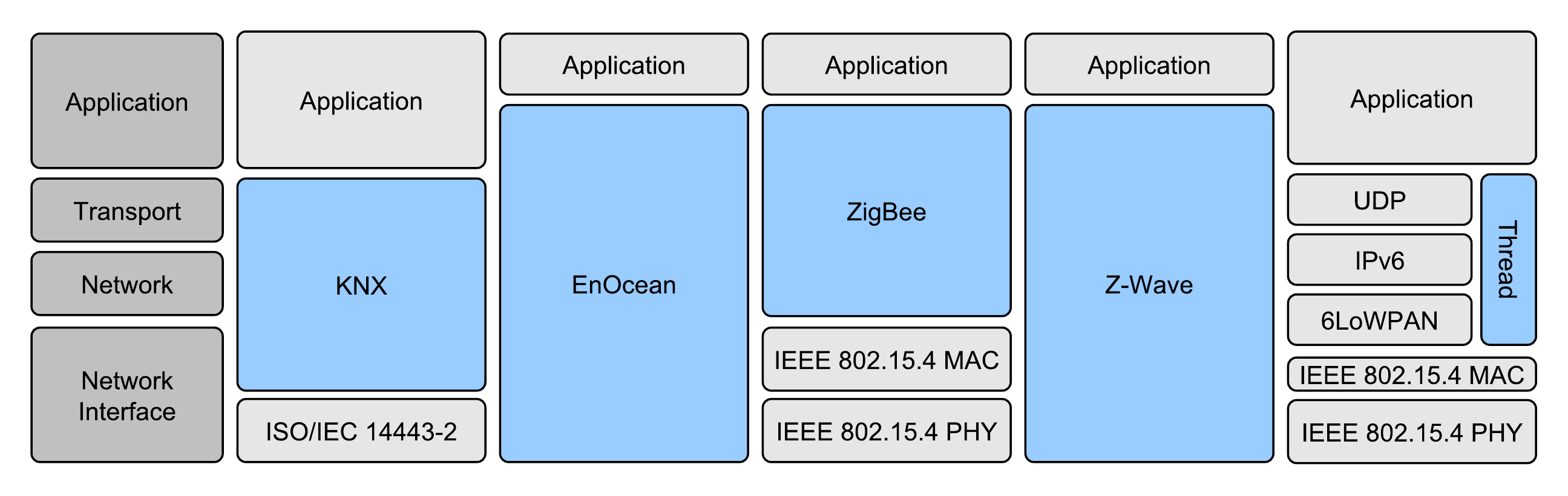}
		\caption{Stack of each protocol according to the TCP/IP Model}
	\label{fig:osi-model-protocols}
\end{figure*}

\section{Protocol Security}
\label{sec:security}

This section contains an analysis of the security measures taken by the protocols from Section \ref{sec:tec}. On
this behalf, it describes the cryptographic methods (algorithms and operation modes) for assuring confidentiality,
authenticity, integrity and replay protection for each protocol.
It further gives an overview of protocol- and implementation-related security issues for each of the protocols.  

\subsection{KNX-RF}
\label{sec:sec:prt:knx}
The KNX protocol itself defines no security measures in KNX, apart from less-than state-of-the-art, plain
text-transmitted passwords \cite{Schwarz:2016}. However, KNX has been extended with \textit{KNX Data Security}, providing encryption, 
authentication and integrity checking using AES-CCM with 128 bits, as well as a sequence number as a
counter. The key exchange is via pre-shared secret, with a project specific tool key derived from a
unique \textit{Factory Device Set up Key (FDSK)}, which is the PSK \cite{knxsec}.  
 Despite some time being in use, a detailed security analysis of
the KNX Security extensions is still not available \cite{glanzer2016increasing}. An earlier analysis of a draft state of
KNXnet/IP Secure, however, found numerous attack vectors and security service shortcomings \cite{6837593}.

\subsection{EnOcean}
\label{sec:sec:prt:EnOcean}
EnOcean provides authentication, integrity checking, encryption and replay protection \cite{EnOcean:Security}.
For encryption, the protocol provides two cryptographic means:
AES-CBC and \textit{variable AES (VAES)}. The EnOcean specification defines for AES-CBC in its schematic description
that ``The initialization vector has all bytes set to 0''.
A constant initialization vector,
however,  is regarded insecure \cite{ENISA:2014}\footnote{An all-zero initialization vector is practice with CBC-MAC,
which, however, is regarded insecure outside the CCM (see Section \ref{sec:sec:prt:zigbee})}, therefore using AES-CBC
in EnOcean is discouraged.
VAES, on the other hand, can be seen as a variant of the \textit{counter mode (CTR)}. The counter in this setup consists
of a specified \textit{Public Key} and a \textit{Rolling Code (RLC)}. The latter should start with a random
number (also called \textit{nonce}) \cite{ENISA:2014}.  
For message authentication and integrity checking, an AES-CMAC is used, which is generally regarded secure under the
conditions that no all-zero payloads occur and that the AES key will change after at most $2^{48}$
messages\cite{ENISA:2014}.
The aforementioned RLC could, apart from being part of VAES, also be a part of a secured EnOcean packet and, in
conjunction with the CMAC, as such be used for replay protection. 
Therefore, when using \textit{EnOcean}, it is recommended to use a \textit{nonce}-starting RLC within the packet for
replay protection, \textit{VAES} for encryption and CMAC for authentication and integrity checking.
All of these (optional) parameters (including a RLC initialization value) are exchanged in a \textit{teach-in mode}.
This mode allows using a pre-shared key, which should be used, as otherwise the exchange will take place
unencrypted, including the session key used later on. 

\subsection{Zigbee}
\label{sec:sec:prt:zigbee}
Zigbee builds on IEEE 802.15.4 (see Section \ref{sec:tec:prt:zigbee}) and 
is therefore subject to security issues concerned with that protocol, which are out of this paper's scope. 
Zigbee's security measures reside, following its architecture, on its network (\textit{NKW}) and application
(\textit{APS}) layers (see Figure \ref{fig:osi-model-protocols}).
The combined encryption and authentication relies on AES-CCM*, which provides both services, and is currently regarded
secure while integrity and replay protection rely, in conjunction with the former, on a message integrity code and a frame
counter\cite{Krausse2014}.
The protocol uses a 802.15.4-derived security level scheme with eight levels (0-7), of which the lower four operate
unencrypted and the upper four are encrypted. For each half, the lowest level (0 and
4, respectively) provides no  authenticity and integrity checking, while the upper ones include increasingly sized
(32, 64, 128 bits of length) \textit{message integrity codes (MIC)} \cite{Zigbee:Spec}. For integrity-checking only (1-3), 
it uses an AES-CBC-MAC, which is discouraged outside of AES-CCM \cite{NIST:2004}. Therefore, only the levels 5-8 can
assure authenticity and integrity.
Zigbee uses two types of keys: \textit{network keys} and \textit{link keys}. The former are mainly
used for broadcasts and capable or securing a Zigbee message at the NKW layer, but have to be known to the entire
network, while the latter may be used for end-to-end security, but only at APS layer. The network
key may be pre-shared or transmitted by a \textit{trust center}, which may have to occur unencrypted, opening a
temporary vulnerability \cite{Krausse2014}. Both versions of message security add auxiliary headers and, if applicable,
a MIC at their respective layer, thus resembling an encapsulating security payload.
The verification of the link (\textit{APS} layer) key is secured using a \textit{Hashed Message Authentication Code
(HMAC)}, where the hash function is a \textit{Matyas-Meyer-Oseas} construction with AES-128 as block
cipher\footnote{Using the same key and block size of 128 bytes thereby eliminates the need for key
compression} \cite{Zigbee:Spec}.

Zigbee can operate in a centralized or distributed manner. The former uses a \textit{trust center} for controlling
security (in particular authorizing new devices and handling key distribution). In distributed Zigbee networks, the
devices of network form a mesh topology, where each router can act as a parent to new devices, while if a network key is
pre-commissioned in some form, there is no additional authorization. Furthermore, application link keys, used for
securing application layer data, are not established in Zigbee relaying that task to a higher level protocol
\cite{Zigbee:Spec}. This lack of specification leaves room for insecure behaviour. 

In the centralized model, a Zigbee network, from a security perspective, forms a star topology with the \textit{trust
center} as a hub. In this setup, despite using strong (cryptographic) building blocks, there are some severe security
flaws, particularly the assumption of key secrecy. 
Essentially, ZigBee provides two methods of key establishment, pre-shared keys and the usage of a publicly
known \textit{Default Trust Center Link Key}, which is common practices for ease-of-interoperability
reasons\footnote{Specified with the value \textit{ZigbeeAlliance09}}. This could compromise initial key exchange
procedures using a built-in fallback mechanism in the standard \cite{Zillner2015}. 
Also, low-cost IoT devices may not
have a secure key storage, so extracting a network key could be a
trivial task. This is even more an issue in the lighting networks using the \textit{Zigbee Light
Link (ZLL)} specification, as ZLL devices should relay on NWK
security\footnote{Besides the fallback, there is also a publicly leaked master key globally securing network key
exchange on all ZLL certified devices.}.
Using the \textit{touchlink} feature, devices can furthermore be forced to associate with a rouge controller from several meters distance \cite{Zillner2015}. 
In this setup, also denial of service attacks, like factory resets and permanent
device disconnects are possible \cite{Morgner:2017:ITA:3098243.3098254}.
Furthermore, a formal analysis of the Zigbee protocol from
2012 yielded indications of vulnerabilities to multiple authentication attack types, specifically timing attacks, that lead to the disclosure of Zigbee's default
key \cite{6178132}.

\subsection{Z-Wave}
\label{sec:sec:prt:zwave}
Z-Wave provides confidentiality, authentication and replay attack protection utilizing a distinct \textit{Security
Layer} within its \textit{Security Command Classes}. Within the latest revision, Zigbee exhibits two classes:
\textit{Security 0 (S0)} for lightweight and \textit{Security 2 (S2)} for stronger security. The latter is divided
into three subclasses: \textit{S2 Access Control}, \textit{S2 Authenticated} and \textit{S2 Unauthenticated}. All of
these classes use AES-128 encryption.
To secure key exchanges, S2 uses \textit{Elliptic Curve Diffie Hellman (ECDH)} schemes and must support
\textit{Curve25519}, with a public key length of 256 bits.  This length conforms with the recommendations of the German
\textit{Federal Office for Information Security (BSI)} \cite{BSI:TR-02102-1}. The curve itself is currently regarded
secure \cite{RFC7748}, although recently a side-channel attack against a specific implementation has been discovered
\cite{Genkin:2017:ITA:3133956.3134029}.
S0 is intended for use with legacy devices, not capable of enforcing S2 security.
In S0, the network key is shared by all devices in a network (similar to Zigbee, see Section
\ref{sec:sec:prt:zigbee}). This is not the case in S2, where each subclass has its own network to prevent a
compromised low-security class device from compromising higher-security class ones. Technically, S2 access control and
authenticated are equivalent, while S2 unauthenticated lacks client authentication capabilities. S2 Encryption and
Authentication are provided by AES-128 in CCM mode, while integrity checking and replay protection rely on an
AES-128-CMAC and \textit{Pre-Agreed Nonces (PAN)} with next-nonce derivation functions.
So far, there are no known security flaws in the protocol itself. However, there are examples of successful examples of
of attacks against specific implementations. For instance, a Z-Wave door lock implementation has been able to be
forced to overwrite its shared network due to the lack of an important state validation check
\cite{fouladi2013security}.

\subsection{Thread}
\label{sec:sec:prt:trd}
As with Zigbee, Thread uses a network-wide key as network-level protection \cite{commthread}. 
This key is used to secure packets via AES-CCM and is exchanged via a variant of
\textit{juggling Password-Authenticated Key Exchange (J-PAKE)}, based on P-256 elliptic curve
Diffie-Hellman\footnote{While there is no clear evidence, there is distrust against NIST elliptic curve standards due
to possible backdoors\cite{Bernstein2016}. P-256 is a NIST-standardized elliptic curve.} 
key exchange (called \textit{EC-JPAKE}).
As this key is known by all devices, the protocol additionally recommends application layer protection.
Therefore, it relies on \textit{Transport Layer Security (TLS)}\cite{RFC5246} and \textit{Datagram Transport Layer
Security}\textit{RFC6347} (both version 1.2).
These protocols provide full security services and are well proliferated. These protocols are also used for the
commissioning and join processes. There are, however, more or less secure configurations, therefore standardized advice
for configuring these protocols should be adhered to \cite{Sheffer2015}.
Using full end-to-end security via TLS or DTLS, however, might have a negative impact on performance and not all small
embedded devices might be capable thereof. As compromising one device might, if TLS or DTLS is not used consequently,
compromise potentially the whole network or at least parts thereof, this circumstance poses a severe security problem.

\section{Conclusion}
\label{sec:conclusion}
\begin{table*}[ht!]
	\centering
	\caption{Security features per protocol}
	\label{fig:tab:feat}
	\resizebox{\textwidth}{!}{
		{\renewcommand{\arraystretch}{1.20}%
		\begin{tabular}{|c|c|c|c|c|c|c|}
			\hline
			\bfseries & \bfseries Authentication/Integrity& \bfseries & \bfseries Encryption& \bfseries & \bfseries & \bfseries \\
			\bfseries & \bfseries Algorithm& \bfseries Length& \bfseries Algorithm& \bfseries Length& \bfseries Replay Protection& \bfseries Key Exchange\\
			\hline 
			Zigbee	&HMAC\_Matyas–Meyer–Oseas	&128	&AES-CCM* (L4-7)	&128	&Counter&PSK or\\ 
			&AES-CBC-MAC (L 1-3)	&	&	&	& &global default\\ 
			&AES-CCM (L 5-7)&	&	&	& &\\ 
			\hline 
			EnOcean	&CMAC\_AES	&128	&AES-CBC	&128	&NONE&PSK or plain\\ 
			&	&	&VAES (AES-CTR)	&128	&Counter&\\ 
			\hline 
			Z-Wave	&CMAC\_AES, AES-CCM	&128	&AES-CCM	&128	&Nonce-based&ECDH\_Curve25519\\ 
			\hline 
			KNX(-RF)&AES-CCM	&128	&AES-CCM	&128	&Counter&PSK\\ 
			\hline 
			Thread	&AES-CCM	&128	&AES-CCM	&128	&Counter&mod. ECDH\_P-256\\
			\hline 
		\end{tabular}
	}
	}
\end{table*}

This paper described the most important IoT protocols for smart home applications in general and more specifically
from a security perspective. From the five considered protocols (Zigbee, EnOcean, Z-Wave, KNX-RF and Thread), all have
encryption and/or authentication/integrity checking services defined, funded on AES with 128 bits (see Table
\ref{fig:tab:feat}), and apart from EnOcean, all of them at support the CCM block cipher
mode for combined encryption and integrity checking. The latter uses a CBC and a Counter mode variant, where the
specification has room for improvement.  
For the Zigbee protocol, there are several
security analyses that revealed blatant vulnerabilities (especially, but not limited to the decentralized mode of operation). Thread is aware of the weakness of sharing a
symmetric network key and tries to compensate this by using TLS-based end-to-end security, which, however, may prove
difficult for large networks in practice and may also be challenging for small embedded devices.  
 Detailed analyses of the finalized KNX security services are still lacking, although an evaluation of a draft version
 of these has revealed a number of security flaws. Z-Wave is also yet not well explored, as it was
 proprietary for a long time before becoming a standard. However, so far only attacks on implementations, not on the
 protocol itself are known. Z-Wave is, however, of the protocols in this paper, the one using the strongest security building
 blocks, utilizing a nonce-based counter and, with S0 and S2 together with its subcategories, a security class-based
 architecture, plus a secure key exchange method based on elliptic curve Diffie-Hellman. Also on Diffie-Hellman relies
 the Thread protocol although on a less trusted curve, while all other protocol rely on less practical or secure method
 as pre-shared or default keys to secure the key exchange.

\section*{Acknowledgment}   
This research was partly funded by the Comet-Project DeSSnet- Dependable, secure and time-aware sensor networks.
DeSSnet is funded in the context of COMET – Competence Centers for Excellent Technologies by BMVIT, BMWFW, Styrian Business Promotion Agency (SFG) and the Province of Styria - Government of Styria, the Carinthian Economic Promotion Fund (KWF) and the Province of Carinthia - Government of Carinthia. The programme COMET is conducted by the Austrian Research Promotion Agency (FFG). The authors are grateful to the institutions funding the DeSSnet project and wish to thank all project partners for their contributions

\bibliographystyle{IEEEtran} 
\bibliography{literature}

\end{document}